\documentclass[A4,prl, twocolumn, superscriptaddress ,showpacs,preprintnumbers,amsmath,amssymb,floatfix]{revtex4}
\usepackage{graphicx}
\usepackage{epstopdf}
\usepackage[usenames]{color}
\definecolor{DarkBlue}{rgb}{0.1,0,0.55}
\definecolor{DarkRed}{rgb}{0.55,0,0.1}
\definecolor{DarkGreen}{rgb}{0.1,0.55,0.1}
\usepackage[colorlinks=true, pdfstartview=FitV, linkcolor=DarkRed, citecolor=DarkBlue, urlcolor=blue]{hyperref}
\usepackage{bm}
\usepackage{subfigure}

\newcommand{\hsys}{H_q}

\newcommand{\bra}[1]{\langle#1\vert}
\newcommand{\ket}[1]{\vert#1\rangle}
\newcommand{\inner}[2]{\bra{#1} #2\rangle}

\DeclareMathOperator{\Tr}{Tr}
\DeclareMathOperator{\sinc}{sinc}

\newcommand{\eqn}[1]{Eq.~(\ref{#1})}
\newcommand{\fig}[1]{Fig.~\ref{#1}}

\begin{document}

\title{
Population inversion of driven two-level systems in a structureless bath
}

\author{T. M. Stace}\email[]{tms29@cam.ac.uk} 
\affiliation{DAMTP, University of Cambridge, Cambridge, CB3 0WA, UK}
\author{A. C. Doherty} 
\affiliation{School of Physical Sciences, University of Queensland,
  Brisbane, Queensland 4072, Australia}
\author{S. D. Barrett}
\affiliation{Hewlett Packard Laboratories,  Filton Road, Stoke Gifford Bristol, BS34 8QZ, UK}


\date{\today}

\begin{abstract}
We derive a master equation for a driven double-dot damped by an unstructured phonon bath, and calculate the spectral density.  We find that   bath mediated photon absorption is important at relatively strong driving, and may even dominate the dynamics, inducing population inversion of the double dot system. 
This phenomenon is consistent with recent experimental observations.
\end{abstract}

\maketitle

Quantum dots are of great current interest due to their controlability, their sensitivity to the quantum nature of the phonon environment \cite{bra99b,agu00,fuj98}, and their potential application to quantum information processing \cite{los98}.  Double-dot systems, formed by depleting a 2D electron gas down to few conduction band electrons in two wells, have shown coherent quantum phenomena \cite{hay03,petta:186802}.

For the last decade, sensitive electrometry has been able to probe double-dot systems in the single electron regime \cite{fie93,aas01}.
Microwave spectroscopy  
reveals peaks when the double-dot is resonant with a driving field, indicating Rabi oscillation \cite{petta:186802,Barrett0412270}.  Similar phenomena have been seen in driven superconducting systems \cite{chi03,lehnert:027002,duty:140503}.

In this paper we study the interaction of a phonon bath with a strongly driven double-dot.  Other work on driven, dissipative two-level systems (2LSs) is reviewed in \cite{gri98}.  We begin by deriving a spin-boson model from microscopic principles, from which we develop a master equation  for the interaction between the double-dot and an unstructured  phonon bath (i.e.\ one lacking resonances).  We analyse the model and find that at zero temperature
and non-resonant driving, phonon assisted photon absorption becomes significant.  
 The rate balance between this and the ever-present phonon induced relaxation results in an increased occupation of the excited state. 

We show that for strong cw-driving and \emph{sub-quadratic} spectral densities,  there are regimes where bath-mediated excitation 
 dominates, leading to the surprising prediction of a large steady-state \emph{population inversion} in a 2LS.  This is consistent with recent experimental results \cite{pettaperscom,petta:186802}.

While inversion is integral to lasers, 
 three levels are usually required.  
{Inversion of nominal 2LSs has
been discussed for superconducting systems \cite{goo04,clerk:176804},
Er-doped glass fibre \cite{Desurvire90} and strongly driven
atom-optical systems \cite{Savage88,Kocharovskaya96,Zhu88}.  These
focus on structured spectral densities or extremely strong
driving in contrast with the mechanism presented here, which only requires significant
coupling to the unstructured bath near the Rabi
frequency. 


\paragraph{Model.}
We consider a periodically driven 2LS (a qubit), modulating both bias and tunnelling, 
\begin{equation}
  \label{eq:model}
  \hsys=-(\epsilon \sigma_z+\Delta\sigma_x )/{2} + {\Omega_0} \cos( \omega_0 t) 
  (\cos\delta\,\sigma_z +\sin \delta \,\sigma_x ),\nonumber
\end{equation}
where $\sigma_z=\ket{l}\bra{l}-\ket{r}\bra{r}$, and $\sigma_x=\ket{l}\bra{r}+\ket{r}\bra{l}$.  In general, $\Omega_0$ is proportional to the slowly varying electric field amplitude and the electric dipole moment.
 The time independent part of $\hsys$ is diagonalised in the energy eigenbasis,  $-\phi \,\sigma_z^{e}/2$, where
$
 \sigma_z^{e}=\ket{g}\bra{g}-\ket{e}\bra{e}=\sin \theta \,\sigma_x + \cos \theta \,\sigma_z$, $\{\ket{g},\ket{e}\}=\{\sin(\theta/2)\ket{r}+\cos(\theta/2)\ket{l},\cos(\theta/2)\ket{r}-\sin(\theta/2)\ket{l}\}$,  $\theta \equiv \arctan(\Delta/\epsilon)$ and  $\phi=\sqrt{\epsilon^2+\Delta^2}$.
We transform $\hsys$ to a frame rotating at the drive frequency according to 
$e^{-i\omega_0 \sigma_z^{e}t/2}$, and make a rotating wave approximation (RWA), discarding terms rotating at frequencies $\phi+\omega_0\gg \eta,\Omega$ 
\begin{equation}
  \label{eq:HRWA}
 \tilde  H_\mathrm{RWA}=-(\eta\, \sigma_z^{e}+
\Omega\,  \sigma_x^{e})/2.
  \end{equation}
  where $\eta=\phi-\omega_0$, $\Omega=\Omega_0\sin \widehat\theta$, $\widehat\theta=\theta-\delta$, and $\tilde{}$ denotes the rotating basis.  This Hamiltonian is diagonalised in the {dressed} basis $\{\ket{-},\ket{+}\}=\{\cos(\varphi/2)\ket{g}-\sin(\varphi/2)\ket{e},\sin(\varphi/2)\ket{g}+\cos(\varphi/2)\ket{e}\}$: $\tilde H_\mathrm{RWA}=-\Omega'\, \sigma_z^{d}/2$ where
$
 \sigma^{d}_z =\ket{-}\bra{-}-\ket{+}\bra{+} =  \sin\varphi \,\sigma_x^{e} + \cos\varphi\,
 \sigma_z^{e}$, $\varphi \equiv\arctan(\Omega/\eta)\in[0,\pi]$ and $\Omega'=({\Omega^2+\eta^2})^{1/2}$ is the effective Rabi frequency.  Note that we work in units where $H$ has dimension $\mathrm{rad/s}$.

We include the influence of phonons, for which the bare Hamiltonian is
$
H_\mathrm{ph}=\sum_\mathbf{q} \omega_\mathbf{q}a^\dagger_\mathbf{q} a_\mathbf{q}
$. The electron--phonon coupling is generically given by \cite{mahan} 
\begin{equation}
H_{\mathrm{e-p}}=\sum_\mathbf{q} i {M_{\mathbf{q}}}\hat\varrho(\mathbf{q})(a_\mathbf{q}-a_\mathbf{-q}^\dagger),
\end{equation}
where $\hat\varrho ({\mathbf{q}})
=\sum_{j,j'}(\int d^3\mathbf{r}\psi_j^*(\mathbf{r})\psi_{j'}(\mathbf{r})e^{-i \mathbf{q}.\mathbf{r}})c_j^\dagger c_{j'}$ is the Fourier transform of the electron density operator and $\{\ket{\psi_j}\}$ form a discrete basis.  $M_\mathbf{q}=C_\mathbf{q}{
(\hbar/2\mu V\omega_\mathbf{q})^{1/2}}$ is the coupling strength, where $\mu$ is the mass density, $V$ is the volume of the lattice and for peizoelectric coupling $C_\mathbf{q}=P$, whilst for deformation coupling, $C_\mathbf{q}=D q$, where $q=|\mathbf{q}|$.  In the two-level approximation the electronic subspace is spanned by the localised states $\{\ket{L},\ket{R}\}$ satisfying $\inner{{\mathbf r}}{L}=\psi({\mathbf r}-{\mathbf d}/2)$ and $\inner{{\mathbf r}}{R}=\psi({\mathbf r}+{\mathbf d}/2)$, where ${\mathbf d}$ is displacement between dots.  Evaluating the integral in $\hat\varrho$, we discard off-diagonal elements such as $M_\mathbf{q}\bra{L}\hat\varrho({\mathbf r})\ket{R}$, since both $|M_\mathbf{q}|\ll\phi$ and $\inner{L}{R}\ll1$ so it is {doubly} small \cite{bra99b,fuj98}. 
 The diagonal elements evaluate to $\bra{L}\hat\varrho({\mathbf q})\ket{L}=e^{-i {\mathbf q}.{\mathbf d}/2}p({\mathbf q})$ and $\bra{R}\hat\varrho({\mathbf q})\ket{R}=e^{i {\mathbf q}.{\mathbf d}/2}p({\mathbf q})$, where $p({\mathbf q})=\int d^3{\mathbf r}|\psi({\mathbf r})|^2 e^{-i {\mathbf q}.{\mathbf r}}$. Therefore 
$\hat\varrho({ \mathbf q})=p({ \mathbf q})(\cos(\mathbf{q}.\mathbf{d} /2)\mathbb{I}-i\sin(\mathbf{q}.\mathbf{d} /2)\sigma_z), 
$ 
 which establishes that phonons couple primarily to the position of the qubit.  For localised states confined to  a region of length $l$, the form factor $p(\mathbf q)\approx1$ for $q\ll1/l$.   The term proportional to $\mathbb{I}$ just perturbs phonon energies, so 
 \begin{equation}
  \label{eq:phonons}
  H_{\mathrm{e-p}} \approx \sigma_z \sum_{\mathbf q} g_{\mathbf q} \left( a_{\mathbf q}^\dagger+a_{\mathbf q} \right),
\end{equation}
where $g_{\mathbf q}=M_{\mathbf q} p({\mathbf q}) \sin(\mathbf{q}.\mathbf{d} /2)/\sqrt{V}$.
The Hamiltonian for the complete system is  $\tilde H_\mathrm{tot}=\tilde H_\mathrm{RWA}+\tilde H_\mathrm{e-p}+\tilde H_\mathrm{ph}$.

We now transform to an interaction picture defined with respect to $\tilde H_\mathrm{free}=\tilde H_\mathrm{RWA}+\tilde H_\mathrm{ph}$,  
in which  $a_{\mathbf q}(t)= a_{\mathbf q} e^{-i \omega_{\mathbf q} t}$ and 
$
  \sigma^{d}_+(t)= \sigma^{d}_+ e^{i \Omega' t}$.  
  Therefore
\begin{eqnarray}
  \sigma_z(t)& =& \left(\cos \theta \cos \varphi -\sin \theta
      \sin \varphi  \cos (\omega_0 t )\right) \sigma^{d}_z \nonumber\\
      &&-\left(\cos
  \theta \sin \varphi + \sin \theta \cos \varphi \cos (\omega_0 t )\right)
  \sigma^{d}_x(t) \nonumber\\
  &&-\sin \theta \sin (\omega_0 t) \sigma^{d}_y(t),\nonumber\\
  &=& \sum_{\omega'\in\{0,\Omega',\omega_0\pm\Omega',\omega_0\} }P_{\omega'} e^{-i \omega' t} +  P_{\omega'}^\dagger e^{i \omega' t}.\label{eq:sigmazt}
  \end{eqnarray}
where $P_0=\cos \theta \cos \varphi\, \sigma^{d}_z /2$,
$  P_{\Omega'} = - \cos
  \theta \sin \varphi\,\sigma^{d}_-$, \mbox{$
  P_{\omega_0 \pm\Omega'} =\mp \sin \theta \left(1\pm\cos \varphi
  \right) \sigma^{d}_\mp /2$, $
P_{\omega_0}= -\sin \theta
  \sin \varphi\, \sigma^{d}_z/2$}.
Then,  the interaction picture Hamiltonian is 
$  H_I(t) = \sigma_z(t)\sum_{\mathbf{q}} g_\mathbf{q} ( 
   a_\mathbf{q}^\dagger e^{i\omega_\mathbf{q}t}+ a_\mathbf{q}
   e^{-i\omega_\mathbf{q} t} )$.

\paragraph{Master equation.} 
To solve the dynamics of the qubit, we develop a master equation for the qubit density matrix, $\rho$. Following \cite{gar00} we integrate the von Neumann equation for the joint density matrix, $W$, of the total system, then trace over phonon modes, resulting in
\begin{equation}
\dot\rho_I(t)=-\int_{t_0}^t dt' \Tr_\mathrm{ph}\{[H_I(t),[H_I(t'),W_I(t')]]\}.\label{eq:vonNeumann}
\end{equation}
We make a Born-Markov approximation, setting $t_0=-\infty$ and replacing $W_I(t')\rightarrow W_I(t)$, which is valid for weak coupling and rapid bath relaxation \cite{gar00}. When $\omega_0$ and $\Omega'$, appearing as rotating terms in $H_I(t)$, are much larger than the phonon coupling strength, we make a second RWA \cite{stace:136802,sta03c}, giving the master equation 
\begin{equation}
  \label{eq:me}
  \dot{\rho_I}=\sum_{\omega'} J(\omega') \left[ \left( N(\omega')+1 \right)
  \mathcal{D} [P_{\omega'}]\rho_I + N(\omega') \mathcal{D} [P_{\omega'}^\dagger ]\rho_I \right],\nonumber
\end{equation}
where $J(\omega)=2\pi\sum_\mathbf{q}|g_{\mathbf q}|^2\delta(\omega-\omega_\mathbf{q})
$ is the spectral density, 
$N(\omega)=(e^{\omega/k_B T}-1)^{-1}$ is the occupation of phonon modes, and $\mathcal{D}[A]\rho\equiv A\rho A^\dagger-(A^\dagger A \rho+\rho A^\dagger A)/2$. 
The different operators $P_\omega$ are classified according to their effect on the system: those proportional to $\sigma_z^d$ produce pure dephasing in the dressed basis, whilst those proportional to $\sigma_\pm^d$ induce transitions between the dressed states. 

The second RWA used to derive the master equation implies that our analysis is valid only in certain limits.  Firstly, the RWA made above is reasonable when $\omega_0,\Omega'\gg J(\omega')$ so we are considering \emph{strong driving} and \emph{weak coupling}, where the Rabi frequency is larger than the dissipation rate.   Weak driving is discussed in \cite{Barrett0412270}.  This implies that on resonance the DC conductance peak is saturated, i.e.\ the average polarisation of the qubit is zero.  
Secondly, the RWA we have made limits analysis of dynamics to times $\gg1/\Omega'$.  In what follows we calculate steady state properties, so the RWA is not restrictive.

\begin{figure}
\includegraphics[height=4cm]{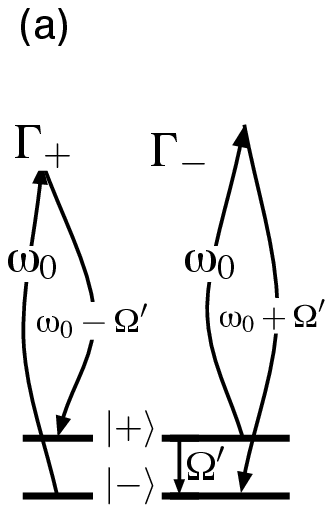}\hspace{1cm}
\includegraphics[height=4cm]{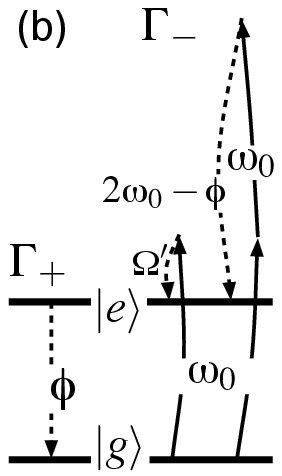}
\caption{Energy level diagrams for dressed state raising and lowering processes in (a) the dressed basis and (b) the bare eigenbasis, for $\eta<0$, where $\ket{-}\approx\ket{e}$ and $\ket{+}\approx\ket{g}$.  Solid arrows indicate photon absorption, and dotted arrows indicate phonon emission.  When $|\eta|\gg\Omega$, $\Omega'\approx|\eta|=\omega_0-\phi$.} 
\label{fig:energylevels}\label{fig:DressedEnergyLevels}\label{fig:BareEnergyLevels}
\end{figure}

Setting
$\rho(t)=(\mathbb{I}+x_d(t)\sigma^d_x+y_d(t)\sigma^d_y+z_d(t)\sigma^d_z)/2,
$
yields the equations of motion for the components of the Bloch vector in the dressed basis,
$
\dot x_d=-(2\Gamma_z +(\Gamma_-+\Gamma_+)/2)x_d$ (and $x_d\rightarrow y_d$), and $\dot z_d= (\Gamma_--\Gamma_+)-(\Gamma_-+\Gamma_+)z_d$, where
\begin{eqnarray}
\Gamma_z&=&(J(0)\cos^2\theta\cos^2\varphi(N(0)+1/2)/2\nonumber\\
&&{}+J(\omega_0)\sin^2\theta \sin^2\varphi(N(\omega_0)+1/2))/2,\nonumber
\\ 
\Gamma_\mp&=&
{J(\omega_0-\Omega')}\sin^2\theta(1-\cos\varphi)^2\frac{ N(\omega_0-\Omega')+\frac{1\mp1}{2}}{\tiny{4}}\nonumber\\&&+{J(\omega_0+\Omega')}\sin^2\theta(1+\cos\varphi)^2\frac{N(\omega_0+\Omega')+\frac{1\pm1}{2}}{4}
\nonumber\\&&+J(\Omega')\cos^2\theta\sin^2\varphi(N(\Omega')+({1\pm1})/{2})
.\label{eq:gamma}
\end{eqnarray}
$\Gamma_z$ contributes to the dephasing rate, but not to transition rates between dressed states.  The three terms appearing in $\Gamma_\mp$ correspond to interactions with phonons of energies  $\omega_0\pm\Omega'$ and $\Omega'$.  For $k_BT\ll\Omega$, $N\approx0$ and the term proportional to $J(\omega_0-\Omega')$ corresponds to a dressed-state raising process, whilst the other two are dressed-state lowering processes, shown in \fig{fig:DressedEnergyLevels}a.
The steady state of the Bloch vector is $x_d=y_d=0$ and $z_d=(\Gamma_--\Gamma_+)/(\Gamma_-+\Gamma_+)$, so $z_d$ is determined by the balance of the dressed-state raising and lowering rates.

We now show that driving can induce  population inversion.  For the purpose of discussion we take $T=0$ and $\epsilon>0$, and we consider the dynamics  below resonance $\eta<0$.  For clarity, we describe the most interesting limit, $|\eta|\gg\Omega$, in which the \emph{lower} dressed state and the bare \emph{excited} state are approximately the same, $\ket{-}\approx\ket{e}$.  In this limit, there is a population inversion 
wherever dressed-state lowering processes dominate,  $\Gamma_->\Gamma_+$. 
 In this limit $\Omega'\approx|\eta|$, $\sin\varphi\approx-R$ and $\cos\varphi\approx-1+R^2/2$, where $R=\Omega/\Omega'$.  Thus $\omega_0-\Omega'\approx\phi$ and $\omega_0+\Omega'\approx 2\omega_0-\phi$ so 
 \begin{eqnarray}
\Gamma_+&\approx&  J(\phi)\sin^2\theta,\nonumber\\
 \Gamma_-&\approx& J(|\eta|)\cos^2\theta\,R^2+J(2\omega_0-\phi)\sin^2\theta\,R^4 /16.  \nonumber
\end{eqnarray}
Typically, $\Gamma_-\ll\Gamma_+$, since the phonon density of states   is small at low energies, so $J(|\eta|)\ll J(\phi)$, and $R^2\ll1$  away from resonance.  However, the condition for inversion, $\Gamma_->\Gamma_+$, can be met in certain regimes, namely when $\sin^2\theta$ is sufficiently small. 

The terms in $\Gamma_\pm$ may be understood physically from \fig{fig:BareEnergyLevels}b, where the dressed-state raising and lowering processes at rates $\Gamma_\pm$ are shown in the undriven eigenbasis.  In this picture, a dressed-state raising process corresponds simply to phonon emission accompanied by \emph{relaxation} of the qubit.  The phonon emission rate contributes the factor $J(\phi)$ to $\Gamma_+$, whilst the factor $\sin^2\theta$ is the dipole matrix element for the phonon-induced transition.  In the absence of driving this is the only significant process \cite{stace:136802}.  

In the presence of driving, dressed-state lowering processes also occur, due to absorption of either one or two photons, accompanied by emission of a phonon of energy $\Omega'\approx|\eta|$ and $2\omega_0-\phi$ respectively.  Both of these processes contribute to $\Gamma_-$, and \emph{excite} the 2LS.  As in the the Jaynes-Cummings model of a driven atom \cite{yamamoto}, away from resonance each photon absorption contributes a factor $R^2$, giving the factors $R^2$ and $R^4$ respectively.  Each photon contribution already implicitly contains a dipole transition matrix element.  Phonon emission contributes factors $J(|\eta|)$ or $J(2\omega_0-\phi)$ respectively.  

Typically the two-photon process is very weak, so inversion occurs when $J(|\eta|)R^2 \cos^2\theta>J(\phi)\sin^2\theta$, i.e.\ when $J(|\eta|)/J(\phi)>\eta^2\tan^2\theta/\Omega^2$. Clearly, this inversion condition is satisfied for sufficiently small $\eta$ if $J$ is sub-quadratic (i.e.\ $J(\omega)<\alpha\,\omega^2$, for some $\alpha$).  These findings are our principle result.  Later we show that inversion does not depend on the limit $|\eta|\gg\Omega$ above, which simply leads to a useful physical interpretation.

The degree of inversion depends on the relative size of $J(|\eta|)$ and $J(\phi)$, as well as the temperature.  As $T$ increases, the phonon modes become thermally occupied, so when $k_b T\gtrsim\Omega'$ the thermally-activated, dressed-state raising process becomes significant.  This is the inverse of the second process depicted in Figs.\ \ref{fig:DressedEnergyLevels}a and \ref{fig:BareEnergyLevels}b, where relaxation of the qubit from $\ket{e}$ to $\ket{g}$ accompanied by stimulated emission of a photon, reducing the inversion.

We now show that the condition $J(\omega)<\alpha\,\omega^2$ is satisfied for piezoelectric phonon coupling (approximately constant or Ohmic,  
depending on phonon dimensionality), and that this is stronger than deformation coupling.  If the spectral density were strongly peaked, e.g.\ due to a discrete phonon spectrum, then the two-photon process may become significant, analogous to \cite{goo04}.

\begin{table}
\begin{center}
\begin{tabular}{|c||c|c|}
\hline
& Deformation & Piezoelectric\\\hline
2D & 
$\frac{\pi^2 D^2}{\mu_2c_s^2d^2}(\frac{\omega}{\omega_p})^2(1-\mathrm{J}_0[2\pi\frac{\omega}{\omega_p}])$ &
$\frac{P^2}{4\mu_2c_s^2}(1-\mathrm{J}_0[2\pi\frac{\omega}{\omega_p}])$\\\hline
3D & 
$\textsf{D}(\frac{\omega}{\omega_p})^3(1-\sinc[2\pi\frac{\omega}{\omega_p}])$ &
$\textsf{P}\frac{\omega}{\omega_p}(1-\sinc[2\pi\frac{\omega}{\omega_p}])$\\\hline
\end{tabular}
\end{center}
\caption{\label{tab:phonon} Spectral density, $J(\omega)$, for deformation and piezoelectric coupling in isotropic 2D and 3D media, assuming $p(\mathbf q)=1$.  $D$ and $P$ are the deformation and piezoelectric constants, $d=|\mathbf{d}|$, $c_s$ is the speed of sound, $\mu_2$ is the areal mass density, $\mu$ is the volumetric density, $\textsf{D}={2\pi^2 D^2\hbar}/{\mu c_s^2d^3}$, $\textsf{P}={P^2\hbar}/{2\mu c_s^2d}$ and $\omega_p=2\pi c_s/d$ is the angular frequency of wavelength-matched phonons.}
\end{table}%

The spectral density $J(\omega)$ varies signicantly over the frequency ranges of interest and depends on both  the dimension of the phonon bath and nature of the coupling to acoustic phonons, either deformation or piezoelectric. 
We assume the phonon bath is either 2 or 3 dimensional, isotropic and dispersionless ($\omega_{\mathbf q}=c_s q$), and the results are summarised in Table \ref{tab:phonon}. Coupling of GaAs double-dots to phonons has been observed in transport measurements \cite{fuj98}, where it was found that spontaneous emission rates are dominated by piezoelectric coupling, with the effective dimension dependent on the qubit splitting.  The angular form factor, $F$, that appears in the 2D spectral densities behaves qualitatively like the corresponding factor $1-\sinc{x}$ in the 3D version. Other effects of phonon coupling to GaAs dots are discussed in \cite{bra99b,bra04}

To quantify the phonon induced decoherence rates, we estimate the characteristic time scales $\tau_\textsf{D}=2\pi\omega_p^3/\textsf{D} \omega_0^3$ and $\tau_\textsf{P}=2\pi\omega_p/\textsf{P} \omega_0$ for 3D phonons, from Table \ref{tab:phonon}.  For GaAs,  $\hbar D=13.7$ eV
, $\hbar P=1.45$ eV/nm, $\mu=5300$ kg/m${}^{3}$, $c_s=5200$ m/s \cite{bru93} and we take $d=500$ nm, $\omega_0/2\pi=24$ GHz \cite{petta:186802}, so $\hbar\textsf{P}=2.3 \mu$eV, $\hbar\textsf{D}=33$ neV and  $\omega_0/\omega_p=2.3$.  Then   $\tau_\mathsf{D}=10$ ns and $\tau_\mathsf{P}=760$ ps, so piezoelectric coupling dominates, in agreement with experiment \cite{fuj98,hay03,petta:186802}.

The variation in the DC response of an electrometer is proportional to  the time-averaged expectation of the double-dot polarisation, $\overline z$,  \cite{stace:136802,sta03c,engel:106804}.  From \eqn{eq:sigmazt} we find  $\overline z=\overline{\langle \sigma_z \rangle}=\cos\theta\cos\varphi\,z_d$.  For weak driving, $\overline z\approx\epsilon/\phi$ away from resonance, whilst on resonance ($\eta=0$) $\overline z=0$, so there is a resonant peak in the electrometer response \cite{petta:186802,Barrett0412270}.  When $|\epsilon|\gg\Delta$, the eigenstates are approximately localised, e.g.\ for $\epsilon\gg\Delta>0$, we have $\ket{g}\approx\ket{l}$, so inversion corresponds closely to localisation in $\ket{r}$.
 
Using the rates in \eqn{eq:gamma}, \fig{fig:SpectralDensity}a shows the  left dot occupation, $M=(1+\overline z)/2$, versus bias, $\epsilon$, for different driving amplitudes, $\Omega_0$, assuming 3D piezoelectric coupling.  We focus on $\epsilon>0$, since the opposite regime is the same (up to reflections).
There is clear population inversion, where $M$ crosses 0.5.  There are several points to note from this figure.  Firstly, the population at resonance always remains at 0.5.  When resonance is approached the dressed states have equal time-averaged projections onto the localised states, as the qubit performs Rabi oscillations through the poles of the Bloch sphere.  Hence the average polarisation vanishes.  Secondly, the peak in the conductance shifts to lower energies, as the photon absorption probability $R^2$ increases with increasing driving.  Both of these features are consistent with recent experimental observations \cite{pettaperscom}.  Finally, the ripples superimposed on the peak are due to the structure of the spectral density, and are spaced by approximately $\omega_p$.  The prominence of the ripples follows from  neglecting anisotropy in the piezoelectric coupling and $c_s$.  
In practise, anisotropy in $c_s$ (as in GaAs) should suppress the ripples.

\begin{figure}
\includegraphics[width=4.2cm]{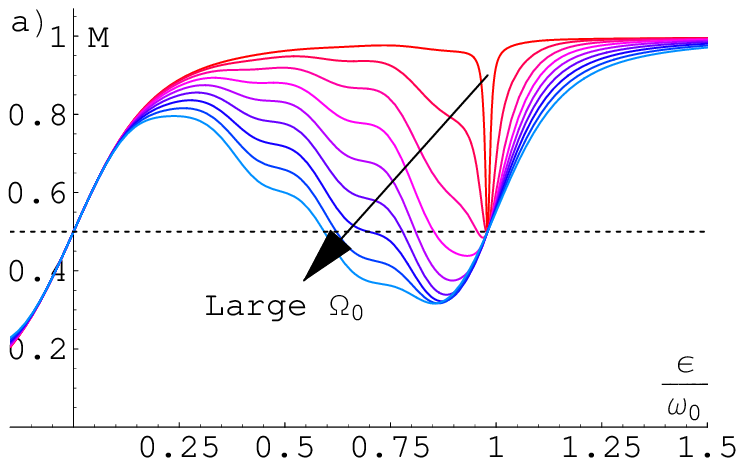}
\includegraphics[width=4.2cm]{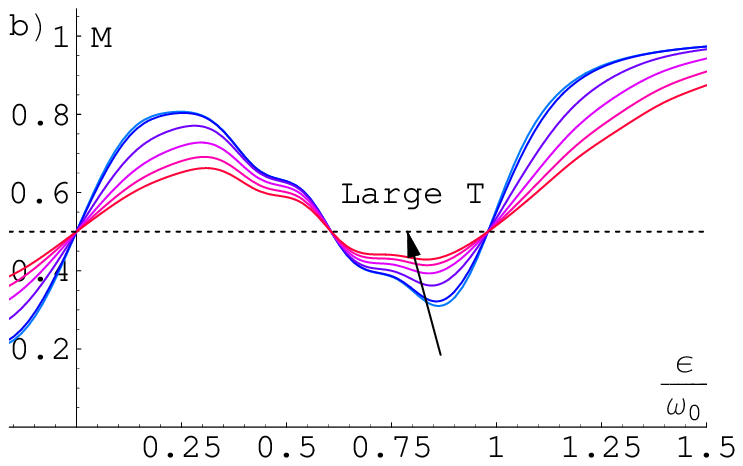}
\caption{
(a) Occupation of left dot, $M$, for increasing Rabi frequency from $\Omega_0=0.01\, \omega_0$ to $0.17\, \omega_0$  in steps of $0.02\,\omega_0$. Other parameters are $\omega_p=0.2\, \omega_0=\Delta$, $k_B T=0.05\,\omega_0$ and $\delta=\pi/2$ corresponding to driving the tunnelling rate. (b) Increasing temperature from $k_B T=0$ to $0.3\,\omega_0$ in steps of $0.06\,\omega_0$.  As above, $\omega_p=0.2\, \omega_0=\Delta$  whilst $\Omega_0=0.16\,\omega_0$.}
\label{fig:SpectralDensity}
\end{figure}

The inversion becomes more pronounced for a 2D piezoelectric coupling, since the spectral density for this case tends towards a constant for  $\omega>\omega_p$.  Transport studies of a similar system  suggest that the actual phonon coupling is intermediate between 2D and 3D \cite{fuj98}, so it is plausible that experimental results will find a more pronounced inversion than that shown above.
  
  In \fig{fig:SpectralDensity}b we show the dependence of $M$ on temperature.  Clearly the resonance is suppressed, and the  average occupation tends to 50\%, as expected.  Also, as $T$ increases, the peak moves towards smaller $\epsilon$.  As discussed above, this occurs because 
when  $k_B T\gtrsim\Omega'$, the thermally-activated dressed-state raising  process (corresponding to relaxation  from $\ket{e}$ to $\ket{g}$) becomes significant, which reduces the inversion.  Since $\Omega'$ increases with detuning, the inversion is more robust further from the resonance condition.  This temperature dependence of the peak, along with its dependence on driving strength, is a signature of the population inversion discussed here.
 
 Since we are considering a regime of relatively strong driving, it is natural to ask whether higher order driving terms are significant.  Such terms produce harmonic resonances, and have been seen experimentally \cite{petta:186802,pettaperscom}. 
There are two origins of harmonic resonances in this kind of system: firstly, at sufficiently high intensity, multi-photon processes may become important.  Secondly, nonlinearity in the coupling of $\Delta$ to the driving voltage, $V(t)$, will produce harmonics of $\omega_0$,  which can drive Rabi oscillations.  Whilst these effects will produce features at $\phi=m\,\omega_0$, unless $\Omega\sim\omega_0$ \cite{goo03}, they will not significantly affect  the system near the fundamental frequency.

In conclusion, we have shown that population inversion is possible for a driven, dissipative 2LS with sub-quadratic spectral density, consistent with recent experimental observations.  The inversion grows as the driving increases, and the maximum inversion moves to lower energies as both driving amplitude and temperature increase.   This phenomenon is quite generic for physically reasonable systems, and does not rely on any specific structure in the spectral density.  

 We thank Jason Petta, Alex Johnson and Charles Marcus for previews of experimental data and discussions, and Mikhail Lukin and Tim Ralph for useful conversations.  
This work was funded by Fujitsu-CMI and the E.U.\ NANOMAGIQC project (IST-2001-33186).

%

Note added: after acceptance at PRL we became aware of
\cite{Dykman79}, which also discusses  population inversion in driven
impurity 2LSs. Although the mechanism discussed in \cite
{Dykman79} is similar to that presented here, we have determined
conditions under which unstructured spectral densities produce inversion, and shown that these may be satisfied in quantum-dot systems. We thank Mark Dykman for bringing \cite{Dykman79} to our attention.
 
%
%

\end{document}